\documentclass[paper]{ieice}
\usepackage[dvips]{graphicx}
\usepackage[fleqn]{amsmath}
\usepackage[varg]{txfonts}
\usepackage{algorithm}
\usepackage{algorithmic}
\usepackage{amssymb}
\usepackage{color}
\usepackage{comment}

\field{}
\title{Cooperative Interference Mitigation Algorithm in Heterogeneous Networks}
\titlenote{This work was supported in part by the National Research Foundation (NRF) of Korea Grant funded by the Korean Government (NRF-2015R1A1A1A05001069), and Institute for Information $\&$ communications technology Promotion (IITP) grant funded by the Korea government (MSIP) (R1010-15-244, Development of 5G mobile communication technologies for hyperconnected smart devices).}
\authorlist{
 \authorentry[vkien@ee.oulu.fi]{Trung Kien VU}{n}{labelAA}
 \authorentry[sungoh@ulsan.ac.kr]{Sungoh KWON}{m}{labelA}[labelC]
 \authorentry[scoh@etri.re.kr]{Sangchul OH}{n}{labelB}
}
\affiliate[labelAA]{The author is with Centre for Wireless Communications, University of Oulu, Oulu, Finland, FI-90014}
\affiliate[labelA]{The author is with the School of Electrical Engineering, University of Ulsan, Ulsan, Korea, 680-749}
\affiliate[labelB]{The author is with the Wireless Application Research Department, ETRI, Daejon, Korea, 305-700}
\paffiliate[labelC]{The corresponding author.}

\begin{document}
\maketitle
\begin{summary}
Heterogeneous hetworks~(HetNets) have been introduced as an emerging technology in order to meet the increasing demand for mobile data. HetNets are a combination of multi-layer networks such as macrocells and small cells. In such networks, users may suffer significant cross-layer interference. To manage this interference, the 3rd Generation Partnership Project~(3GPP) has introduced enhanced Inter-Cell Interference Coordination (eICIC) techniques. Almost Blank SubFrame (ABSF) is one of the time-domain techniques used in eICIC solutions. We propose a dynamically optimal Signal-to-Interference-and-Noise Ratio~(SINR)-based ABSF framework to ensure macro user performance while maintaining small user performance. We also study cooperative mechanisms to help small cells collaborate efficiently in order to reduce mutual interference. Simulations show that our proposed scheme achieves good performance and outperforms the existing ABSF frameworks.
\end{summary}
\begin{keywords}
eICIC, ABSF, Interference Mitigation, Small Cell, HetNets.
\end{keywords}

\section{Introduction}
\label{Intro}
\begin{figure}[ht]
  \centering
  \includegraphics[scale=0.4]{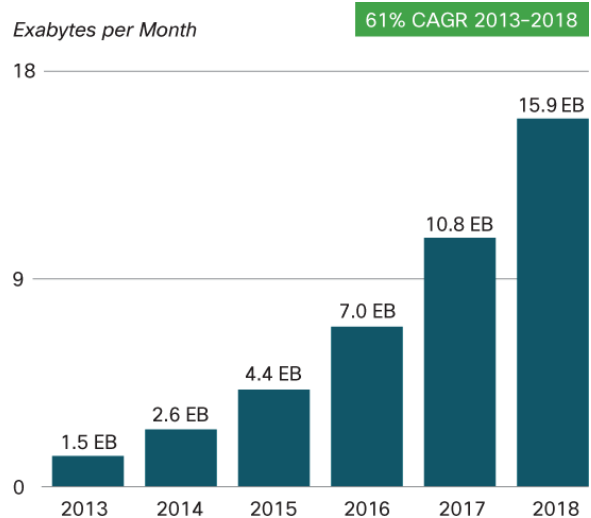}\\
  \caption{Cisco forecasts 15.9 exabytes per month of mobile data traffic by 2018~\cite{Cisco2013}}\label{lb-MobileData}
\end{figure}

In wireless system development, the demand for massive data traffic has seen exponential growth due to the exponential increase in the number of mobile broadband subscribers using smart phones, tablets, and other media devices~\cite{Cisco2013}. As can be seen in Fig.~\ref{lb-MobileData}, the overall mobile traffic is forecast to reach 15.9 exabytes per month by 2018, nearly 11 times the traffic seen in 2013. Small cells are emerging techniques to meet the above challenges by incorporating spectrum sharing, and additional cells in order to provide indoor and outdoor wireless services and recover cell edge user performance and offload macrocells~\cite{D2011survey}.

Small cells are deployed under the coverage of a macrocell network, referred to as heterogeneous networks~(HetNets). A HetNet was introduced as a flexible, low-cost, energy-efficient solution by the 3rd Generation Partnership Project~(3GPP) Long-Term Evolution-Advanced (LTE-A)~\cite{D2011survey,Khandekar2012,TranSS2012}. A HetNet consists of multiple types of access nodes in wireless networks such as macrocells and small cells. According to the capacity of cells and the objectives of deployment, small cells are categorized into three kinds of small cells: microcells, picocells, and femtocells. Micro and picocells are used to expand the coverage and capacity in some specific areas such as airports, shopping malls, and hospitals, whereas femtocells are solutions for wireless data services in indoor environments (e.g., apartments and offices). Femtocells are denoted as Home evolved NodeBs~(HeNBs) for LTE.

In contrast to HetNets, traditional wireless cellular networks consist of only macrocells to serve all users and are called homogeneous networks. In such networks, macrocells have homogeneous characteristics such as transmission power, antenna patterns, noise parameters, propagation models, and connect with each other through similar backhaul connectivity. The users under the coverage of a serving macrocell may experience high interference from neighboring macrocells. To reduce inter-cell interference, the deployment of macrocells has been studied; optimizing macrocell configuration can maximize network coverage and capacity and limit interference. Due to the requirements imposed by high levels of mobile data traffic, the number of macrocells to be deployed increases, which raises the cost of network services and makes deployment difficult in some areas.

In a HetNet, macrocells and small cells share the same radio frequency spectrum, since the reuse of resources can help to increase network capacity and reach the peak data rate. Femtocells prefer to work in Closed Subscriber Group (CSG) mode that allows a limited number of registered small cell users to connect. This configuration creates coverage holes inside the macrocells, called \emph{black~holes}, where the macro users located within the transmission range of the HeNBs cannot be served by either HeNBs or the macrocell, since the macro users experience high Inter-Cell Interference (ICI)~\cite{zahir2013i}. Beside the CSG access mode, open and hybrid access modes are also introduced with more access options such as open access mode, wherein all users (macro users and small cell users) located within the transmission range of the HeNBs can connect to data services, and hybrid access mode, which allows some specific unregistered macro users and registered small cell users to connect. However, the open access mode is designed for public services rather than for private services, which are deployed by the end user who buys the devices and pays the backhaul and electricity. The hybrid mode solution seems to be acceptable, but when the number of unregistered macro users increases, the performance of registered small cell users is degraded.  Further discussions of access mode can be found in~\cite{de2010a,Kwon2011,Kwon2013}. In the near future, hyper-dense deployment of small cell networks is expected to be  realized, but because of co-channel deployment and CSG access mode, interference management is a crucial challenge  that must be addressed in order to obtain the most benefit from small cell networks.

To cope with such an ICI problem, enhanced Inter-Cell Interference Coordination (eICIC) techniques were proposed in Release 10~\cite{R1-104256,R1-104968,R1-104661}. The eICIC solutions include some techniques such as time-domain, frequency-domain, and power control techniques. Almost Blank Sub-Frame (ABSF) is a time-domain technique in which the interfering layer will stop using some time-domain resources, called muted time-domain resources, allowing the victims scheduled use of these resources in order to get rid of ICI. The technique is called almost blank, because some sub-frames cannot be muted at all to transmit the control signals such as Common Reference Signals~(CRS). The ABSF technique requires accurate time synchronization among all cells. The ABSF technique can be used in a different way for both macro-pico and macro-femtocell networks. In macro-picocell networks, the macrocell layer is considered the interfering tier, and the victim macro users are allowed to connect to the picocell instead of the macrocell~\cite{Wang2012,Pedersen12,David2011b}, whereas in the macro-femtocell networks, since femtocells in CSG access mode  become the interfering tier, the femtocells stop using some time-domain resources in such networks~\cite{David2011a}.

The ABSF technique is a simple and efficient solution; however, how many and which time-domain resources need to be muted should be carefully considered before applying the ABSF technique in order to reduce cross-tier interference between the macrocell and small cell layers. In recent years, the eICIC technique has been studied for both macro-picocell and macro-femtocell networks~\cite{David2011a,MIKI2012}. In~\cite{Pedersen12}, the authors explain the benefits and characteristics of the eICIC technique. In a macro-picocell network, the Cell Range Extension (CRE) is used in picocells to extend the network coverage of picocells, such that the users can access picocell networks. However, these users may suffer interference from macrocells. The ABSF technique is now used in the macrocell by muting some macrocell time-domain resources in order to help picocells serve their users with acceptable interference.

In~\cite{Wang2012,Madan2010}, the recommended settings of the CRE offset and the muting ratio in different macro-pico scenarios were proposed. In~\cite{Jiang2012}, the authors proposed a resource allocation scheme for eICIC in macro-pico networks using ABSF selection and a UE partition scheme, which both seek to maximize the network utility function without considering the Quality of Service~(QoS) requirement of users. The ABSF selection and the UE partition scheme are selected in turn to achieve performance balance among the cells. A fixed ABSF pattern is set under each UE partitioning, and a new ABSF ratio is selected for all eNBs for given UE scheduling.

In a macro-femtocell network scenario, the ABSF framework is proposed to track the macro users and an estimation of the Signal-to-Interference-and-Noise Ratio~(SINR) is presented to trigger the femtocells and activate the ABSF mode in~\cite{Kamel2012}. The performance gain of the ABSF framework with a fixed ratio showed a significant improvement in macro user throughput as compared to the non-ABSF technique. In~\cite{Kamel2013}, the problem with ABSF ratio selection for normal macro users and victims is also modeled as a network utility maximization without considering the QoS requirements, which depend on the number of victim macro users and the total number of macro users. In addition, the proposed ABSF also showed that the ABSF ratio for each femtocell depends on the number of nearby victim macro users. The coalition formulation scheme in~\cite{Kamel2013} grouped the VMUEs and forced each aggressor femtocell to stop transmission at different starting subframes. However, this coalition configuration posed the following problem: if two or more adjacent aggressor femtocells affecting the same victim macro users stop transmitting data at different subframes, the victim macro users are still affected by one of the aggressors during these subframes.

Moreover,~\cite{Lembo2013,Cierny13} derived the necessary number of ABSFs based on network parameters such as the number of non-victim macro users and femtocell users that leads to imbalance in the performance of users. The same muted rate is set globally for all small cells, this configuration is not realistic or optimal in practical environments. Since the nature of unplanned and random deployment of small cells makes each small cell unique with respect to interference impact, the number of ABSFs for small cells should be set differently and locally based on the radio resource demand of the nearest vulnerable users in order to avoid inefficient use of radio resources.

In our work, we propose an efficient ABSF framework in order to mitigate the downlink interference for the macro-femtocell networks. Unlike the previous work, we propose a framework to derive the number of ABSFs based on the QoS of victim macro users. Although the macro users are more primary than the small cell users, an efficient ABSF framework offers a necessary and sufficient amount of radio resources for the macro users without too great of an effect on the performance of small cell users. The ABSF computation method is implemented in various environments, such as static to dynamic and sparse to dense network environments. We also propose a cooperation mechanism in order to help aggressor HeNBs in adjacent network areas cooperate to use network resources efficiently, which solves the the coalition problem in~\cite{Kamel2013}.

The rest of this paper is organized as follows. Section~\ref{lb-SM-P} describes the system model and problem. Section~\ref{ProposedAL} introduces an algorithm for setting the number of ABSFs required for MUEs and HeNBs and a framework for cooperation among aggressor HeNBs. In Section~\ref{Evaluation}, we present simulation results to demonstrate the performance of our proposal. We conclude the paper in Section~\ref{Conclusion}.

\section{System Model and Problem}
\label{lb-SM-P}
\subsection{System Model}
\label{lb-SystemModel}
We consider the downlink (DL) of HetNets as depicted in Fig.~\ref{lb-Sys}. The macro evolved NodeB~(macro eNB) is located at the center of each cell and the macro user equipments~(MUEs) are randomly deployed within the coverage of the macro eNB. The HeNBs are randomly deployed by the end user, and each one serves one home user equipment (HUE).

In HetNets, each cell consists of a macro eNB serving $\emph{M}$ MUEs, and let $\mathcal{M} = \{1,..., \emph{M} \}$ denote the set of MUEs. $\emph{F}$ small cells HeNBs are randomly located within the coverage of the  macro eNB, and let $\mathcal{F} = \{1,..., \emph{F} \}$ denote the set of HeNBs; each HeNB $f \in \mathcal{F}$ works in CSG access mode that allows only one HUE to connect. Let $\mathcal{L}$ denote a set of links from the macro eNB to its serving MUEs, $\mathcal{L} = \{ \emph{l}_{1}, ..., \emph{l}_{\emph{M}} \}$, where $\emph{l}_{m}$ is the communication link between the  macro eNB and MUE $m, m \in \mathcal{M}$.

The X2 interface is assumed to exchange signal information among the cells. The functionality of the X2 interface can be used to do some tasks such as load management, inter-cell interference coordination, handover cancellation, and error handling~\cite{X22011}.
\begin{figure}[t]
  \centering
  \caption{System Model}\label{lb-Sys}
\end{figure}

We assume that the HeNBs work in CSG access mode, in which the HeNBs do not allow the MUEs to connect. If an MUE served by the macro eNB is located around a nearby HeNB, it experiences high interference from the nearby HeNB. We define an MUE that affected by the nearby HeNB as a victim macro user~(VMUE), and that HeNB is called an interfering HeNB or an aggressor HeNB.

In order to determine the VMUEs, the SINR is considered a measure of the quality of the links. The SINR can be defined as the received power from the serving cell divided by the sum of the interfering power (from all the other interfering signals) and the noise power. Since the interfering signals from other neighboring cells can affect link quality, the SINR is an important metric that takes into account interference in order to detect VMUEs.

The SINR $ \gamma_{m} $ at link $\emph{l}_\emph{m} \in \mathcal{L}$ between the  macro eNB and the MUE $ \emph{m} \in \mathcal{M}$ is defined as
 \begin{eqnarray} \label{eq:sinr-m}
  \gamma_{m} &=& \displaystyle \frac{G_{(M, m)} P}{\Sigma _{f \in \mathcal{F}} G_{(f, m)} p + \sigma_{m}}, \nonumber \\
             &=& \displaystyle \frac{G_{(M, m)} P}{\eta_{m}}.
 \end{eqnarray}
where $\sigma_{m}$ is the thermal noise at the MUE $m$. Let $P$ and $p$ denote the transmission power of macro eNB and HeNB, respectively, $G_{(M, m)}$ and $G_{(f, m)}$ are the path gains between the  macro eNB and the MUE $m$ and between the HeNB $\emph{f}$ ($ \emph{f} \in \mathcal{F}$) and the MUE $m$, respectively, and $\eta_{m}$ is the sum of interference and noise at the MUE $m$.
\begin{figure}[t]
  \centering
  \includegraphics[scale=0.5]{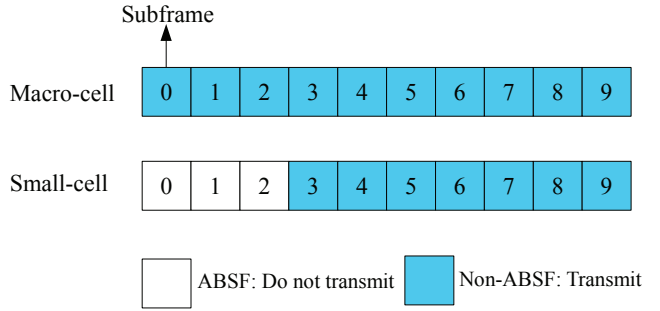}
  \caption{The almost blank subframes}\label{lb-absf}
\end{figure}

As illustrated in Fig.~\ref{lb-absf}, an HeNB mutes a part of the subframes. The blanking rate in the HeNB helps to improve MUE performance, but affects HUE performance degradation. Since the HeNB is located at different places in the network coverage, it has unique characteristics, as illustrated in~Fig.\ref{lb-Sys}. Hence, the ABSF for each HeNB should be optimal and different in order to utilize the network resources effectively. We propose an approach to dynamically select the optimal ABSF ratio based on QoS required for victim MUEs in each aggressor HeNB. In addition, we divide small cells into groups in order to help aggressor HeNBs cooperate.

\subsection{Problem Formulation}
\label{lb-Obj}
Let $\alpha_{m}$ be the muted rate required for a MUE $m$ in order to obtain the radio resources needed to get rid of high interference, $\alpha_{m}$ can be defined as the ratio of the number of muted subframes to the total number of all subframes in a single frame. Our objective is to determine the minimum $\alpha_{m}$ that satisfies the SINR $\gamma_{m}$ of the MUE $m$, which can be expressed as
\begin{equation}\label{eq:Obj}
\begin{aligned}
& {\text{minimize}}
& & \sum_{m \in \mathcal{M}} \alpha_{m} \\
& \text{subject to}
& &  \gamma_{m} \geq \gamma_{m}^{min}, m \in \mathcal{M}, \\
& & & 0 \leq \alpha_{m} \leq 1.
\end{aligned}
\end{equation}
where $\gamma_{m}^{min}$ is the SINR threshold of the MUE $m$. The value of $\gamma_{m}^{min}$ for each MUE $m$ is different, since the MUEs can be affected by different HeNBs. In order to detect whether there are victim MUEs within the coverage of the HeNBs, two approaches are proposed to determine the presence of victim MUEs based on the reported MUE measurements and the basis of detection of uplink transmission from victim MUEs~\cite{TR-36921}. Other approaches are designed to determine the existence of indoor victim MUEs~\cite{Yang2011}. In this paper, the main focus is to determine the presence of nearby outdoor victim MUEs.

Based on the muted rate $\alpha_{m}$ of the victim MUEs, the number of subframes in which the aggressor HeNBs should be muted can be set, and thus the HeNBs cooperate in order to work in ABSF mode efficiently.

\section{Proposed Algorithm}
\label{ProposedAL}
In this section, our proposed algorithm aims to select an appropriate ABSF ratio based on the SINR for macro-femtocell networks in order to reduce the dominant cross-tier interference. In Fig.~\ref{lb-Sys}, MUEs~1,~2, and~3 receive a strong interference signal from nearby HeNBs~1,~2, and~3, respectively. By using the ABSF technique in an HeNB, the HeNB configures some subframes as almost blank subframes. This means that, if during these subframes data transmission with the HeNB stops, then the MUE can schedule its transmission to overcome the interference. However, this configuration degrades HUE performance. Hence, the ABSF ratio in the aggressor HeNBs should be selected carefully to maintain the performance of the HUEs while satisfying the required performance of the MUEs. The proposed algorithm flows can be understood as follows: first, the SINR-based ABSF ratio required by the VMUE is estimated; then, based on the same VMUE, the aggressor HeNBs are grouped as a coalition in order to determine their own ABSF ratio and to cooperate to configure their muted subframes efficiently. We describe the overall proposed algorithm based on the assumption that all tasks are performed by the macro eNB. For the distributed algorithm, more discussion will be provided later.

The proposed algorithm performs based on the following steps:
\begin{itemize}
  \item \emph{Step 1}: MUEs report to the  macro eNB information such as the received power, aggressor HeNBs, and then the macro eNB determines the victim MUEs.
  \item \emph{Step 2}: The macro eNB computes the required ABSF ratio for each victim MUE. The ABSF ratio of the VMUEs is calculated based on their SINR requirements via the ABSF calculation mechanism as will be described later in subsection~\ref{ABSF-Sel}.
  \item \emph{Step 3}: The macro eNB collects the VMUEs that are affected by each aggressor HeNB based on the information reported by the VMUEs and finds the minimum muted rate to satisfy all victim MUEs in each aggressor HeNB as described later in subsection~\ref{Femto-Collect}.
  \item \emph{Step 4}: The coalition algorithm is executed to group the aggressor HeNBs by the  macro eNB in subsection~\ref{Femto-Group}. The strategy of the coalition algorithm is that if two or more aggressor HeNBs have the same VMUEs, they will be joined into a coalition.
  \item \emph{Step 5}: The aggressor HeNBs operate the ABSF mode.
\end{itemize}

To select the ABSF ratio, collect the VMUEs, and group the aggressor HeNBs, we propose the three following mechanisms: ABSF computation, the VMUE collecting algorithm, and the aggressor HeNB coalition, respectively.

\subsection{ABSF Computation for Aggressor HeNBs}
\label{ABSF-Sel}
For~\emph{Step 2}, we propose an ABSF calculation mechanism in order to select the appropriate muted rate required for the VMUEs. The previous work~\cite{Lembo2013,Cierny13} derived the optimal ABSF ratio based on the number of users, including macro users and femtocell users, but this configuration may not guarantee the network performance requirement. In this paper, the optimal ABSF ratio is derived based on the QoS of all the MUEs in order to satisfy the MUE performance requirement, but limit the effect on the HUE performance requirement. Each VMUE may request a different muted rate in order to overcome the interference that affects its data transmission.

To ensure the quality of the signal from a macro eNB to each MUE, each link $m$ has a minimum requirement in terms of SINR~\cite{Kwon2009}, i.e $\gamma_{m} \geq \gamma_{m}^{min}$. By using~(\ref{eq:sinr-m}), the constraint~(\ref{eq:Obj}) can be rewritten in matrix form as
  \begin{equation} \label{eq:Obj-matrix}
  \mathbf{F} \mathbf{P}_{F} \leq  \mathbf{P}_{M} - \mathbf{b},
 \end{equation}
 where the power vector of macrocell $\mathbf{P}_{M} = (P_1, ..., P_M)^T$, the power vector of femtocell $\mathbf{P}_{F} = (p_1, ..., p_M )^T$, $\mathbf{b} = (b_1, ..., b_M)^T$ such that $b_m = \displaystyle \frac{\gamma_{m}^{min} \sigma_{m}}{G_{(M, m)}}$, ($m \in \mathcal{M}$), and $\mathbf{F}$ is a non-square matrix with $M \times F$ elements that can be defined as
\begin{equation} \label{eq:sirn-matrixF}
\mathbf{F}(m,f) = \displaystyle \frac{ G_{(f, m)} \gamma_{m}^{min} }{ G_{(M, m)} }.
\end{equation}

If the value of element $\mathbf{F}(m,f)$ is non-zero, this means that HeNB $f$ creates downlink interference to the corresponding MUE $m$.

When the HeNB is marked as an aggressor HeNB, it will stop using some subframes to limit its interference to the victim MUE. Then the SINR $\gamma_{m}$ at link $\emph{l}_\emph{m}$ between the  macro eNB and the MUE $m$~(\ref{eq:sinr-m}) can be rewritten as
 \begin{eqnarray} \label{eq:sinr-m-alpha}
  \gamma_{m} &=& \displaystyle \frac{G_{(M, m)} P}{\Sigma _{f \in \mathcal{F}} G_{(f, m)} p (1 - \alpha_{m}) + \sigma_{m}}.
 \end{eqnarray}

Now, our objective~(\ref{eq:Obj-matrix}) can be transformed as
\begin{equation}\label{eq:Obj-Sol}
\begin{aligned}
& {\text{minimize}}
& & \sum_{m \in \mathcal{M}} \alpha_{m}  \\
& \text{subject to}
& &  \mathbf{A} \alpha \succeq \mathbf{B}, \\
& & & 0 \leq \alpha_{m} \leq 1.
\end{aligned}
\end{equation}
where $\mathbf{A} = \mathbf{F} \mathbf{P}_{F}$, $\alpha = \{\alpha_1, ...,\alpha_M \}$, and $\mathbf{B} = \mathbf{b} + \mathbf{F} \mathbf{P}_{F} - \mathbf{P}_{M} $. The operator $\succeq$ is for element-wise comparison of two matrixes. Since $\mathbf{A}$ is a non-square matrix, a solution~\cite{chong2013} to problem $\mathbf{A} \alpha_{m} = \mathbf{B}$ that minimizes $\alpha_{m}$ is
\begin{equation}\label{eq:solution}
 \alpha_{m}^{\ast} =  \mathbf{A}^{T} (\mathbf{A} \mathbf{A}^{T})^{-1} \mathbf{B}.
\end{equation}

\subsection{Algorithm for VMUE Collection}
\label{Femto-Collect}
In densely deployed femtocell networks, the number of macro users and HeNBs becomes extremely large and uncontrollable. In order to reduce the dominant cross-tier interference in this scenario, the neighboring HeNBs should form a coalition in order to mute during the proper time-domain subframes. As illustrated in~Fig.\ref{lb-Sys}, MUE~2 is affected by two neighboring HeNBs $2$ and $3$; if HeNB~$2$ mutes at time slot $t_{1}$ and HeNB~$3$ mutes at time slot $t_{2}$, $( t_{1} \neq t_{2} )$, then during time slot $t_{1}$ and $t_{2}$ MUE~2 still affected by either HeNB~$3$ or HeNB~$2$. Hence, if a VMUE is affected by multiple HeNBs in the coalition, the group of these HeNBs has to stop data transmission at the same time.

We propose a cooperative method that enables aggressor HeNBs to collaborate based on the same VMUEs by grouping formation, thus mitigating the downlink interference from two or more nearby aggressor HeNBs to the same VMUEs. The main idea of this method is to group aggressor HeNBs that create downlink interference on the same VMUEs.

For the proposed algorithm, coalition formulation is performed by the macro eNB in the case of a high computing system and energy supply. The coalition formulation can be done via application of two algorithms: the VMUE collecting algorithm (Algorithm~\ref{A-I}) and the  mutual aggressor HeNB grouping algorithm (Algorithm~\ref{A-II}). In Algorithm~\ref{A-I}, the MUE will report its status (including its received and noise power levels and aggressor HeNBs) to its serving  macro eNB, the  macro eNB then determines whether the MUE is a VMUE or not and what aggressor HeNBs are the main aggressors to that VMUE. Algorithm~\ref{A-I} is then performed in order to list the subset of VMUEs that are affected by each aggressor HeNB. The operation of Algorithm~\ref{A-I} is based on a loop such that at the beginning, the first aggressor HeNB checks its presence in a set of aggressor HeNBs for each VMUE $v_{n}$; if it appears, the aggressor HeNB inserts the VMUE $v_{n}$ into the affected list. The loop will continue until the last aggressor HeNB.
\begin{algorithm}[ht]
\caption{The VMUE Collecting Algorithm}
\label{A-I}
\begin{algorithmic}
\STATE $\mathcal{V}$: the set of VMUEs $\mathcal{V} = \{v_{1}, v_{2}, ...,v_{N}\}$, where $N$ is the number of VMUEs
\STATE $\mathcal{A}_{v_{n}}$: the set of aggressor HeNBs affecting VMUEs, $v_{n}$
\STATE Find the set of VMUEs $\mathcal{V}_{f}$ affected by an aggressor HeNB $f$

\FOR{$f = 1$ to $F$}
\STATE $\mathcal{V}_{f} = \O$
\FOR{$n = 1$ to $N$}
\IF{ $ \{ f \} \cap \mathcal{A}_{v_{n}} \neq \O$}
\STATE $\mathcal{V}_{f} = \mathcal{V}_{f} \cup \{v_{n}\}$
\ENDIF
\ENDFOR
\ENDFOR
\end{algorithmic}
\end{algorithm}

We explain the VMUE collecting algorithm (Algorithm~\ref{A-I}) by offering the following example. As can be seen in Fig.~\ref{lb-Sys}, the set of VMUEs $\mathcal{V}$ includes $\{MUE~1,~MUE ~2,~MUE ~3,~MUE~6\}$, the set of aggressor HeNBs affecting $MUE~1$ is $\mathcal{A}_{v_{1}} = \{HeNB~1\}$, the set of aggressor HeNBs affecting $MUE ~2$, $\mathcal{A}_{v_{2}} = \{HeNB ~2, HeNB~3\}$, the set of aggressor HeNBs affecting $MUE ~3$ is $\mathcal{A}_{v_{3}} = \{ HeNB ~3 \}$, and  the set of aggressor HeNBs affecting $MUE ~6$ is $\mathcal{A}_{v_{6}} = \{ HeNB ~4,~HeNB ~5,~HeNB ~6 \}$. Since the aggressor $HeNB ~1$ finds itself in the list $\mathcal{A}_{v_{1}}$, then the $MUE ~1$ will be attached in the set of VMUEs affected by an aggressor $HeNB ~1$, $\mathcal{V}_{1} = \{ MUE ~1 \}$. Similarly, we obtain $\mathcal{V}_{2} = \{ MUE ~2,  MUE ~3 \}$, $\mathcal{V}_{3} = \{ MUE ~3 \}$, and $\mathcal{V}_{4} = \mathcal{V}_{5} = \mathcal{V}_{6} = \{ MUE ~6 \}$.

\subsection{Aggressor HeNBs Coalition Algorithm}
\label{Femto-Group}
To group the mutual aggressor HeNBs into a coalition based on the victim MUE lists from the Algorithm~\ref{A-I}, Algorithm~\ref{A-II} uses a set intersection algorithm in order to find the mutual aggressor HeNBs to the same VMUEs. This process is done by the  macro eNB for the centralized approach. The main idea of Algorithm~\ref{A-II} is that if any aggressor HeNB has the same VMUEs as another, they will form a coalition. Otherwise, one aggressor HeNB can create a separate coalition. The loop starts over at the last aggressor HeNB, and if an aggressor HeNB sees that any other aggressor HeNB has the mutual VMUEs, then they will form a coalition.
\begin{algorithm}[ht]
\caption{The Mutual Aggressor HeNB Grouping Algorithm}
\label{A-II}
\begin{algorithmic}
\STATE $\mathcal{V}_{f}$: the set of VMUEs affected by an aggressor HeNB $f$
\STATE ${flag}_{f}$: a flag associated with $\mathcal{V}_{f}$. if $\mathcal{V}_{f}$ is grouped, ${flag}_{f} = 1$, if not ${flag}_{f} = 0$
\STATE $\mathcal{C}_{c}$: the group of mutual aggressor HeNBs, where $c$ is the index of the group
\STATE $\mathcal{G}_{c}$: the coalition of VMUEs affected by a group of mutual aggressor HeNBs 

\FOR{$c = 1$, $f = 1$ to $F$}
\IF{${flag}_{f} = 0$}
\STATE $\mathcal{G}_{c} = \mathcal{V}_{f}$, $\mathcal{C}_{c} = \{ f \}$, ${flag}_{1} = 1$
\FOR{$f = 1$ to $F$}
\IF{$\mathcal{G}_{c} \cap \mathcal{V}_{f} \neq \O$ and ${flag}_{f} = 0$}
\STATE $\mathcal{G}_{c} = \mathcal{G}_{c} \cup \mathcal{V}_{f}$, $\mathcal{C}_{c} = \mathcal{C}_{c} \cup \{ f \}$, ${flag}_{f} = 1$
\ENDIF
\ENDFOR
\STATE $c = c + 1$
\ENDIF
\ENDFOR
\end{algorithmic}
\end{algorithm}

An example is displayed in Fig.~\ref{lb-Sys-Coa}; there are three mutual aggressor HeNBs groups such that $\mathcal{C}_{1} =  \{ HeNB~1\}$, $\mathcal{C}_{2} =  \{ HeNB~2,~HeNB~3\}$, and $\mathcal{C}_{3} =  \{HeNB~4,~HeNB~5, HeNB~6\}$ with three coalitions of VMUEs affected by each group $\mathcal{G}_{1} = \{ MUE~1 \}$, $\mathcal{G}_{2} = \{ MUE~2,~MUE~3 \}$, $\mathcal{G}_{3} = \{MUE~6\}$, respectively.
\begin{figure}[t]
  \centering
  \caption{An example of coalitions}\label{lb-Sys-Coa}
\end{figure}

Each HeNB in each coalition determines its muted rate based on the muted rate of the VMUEs belonging to its coalition. The aggressor HeNB may work with different muted rates for different MUEs. Hence, to satisfy all victim MUEs belonging to the aggressor HeNB $f$, the muted rate for the HeNB $f$ is $\alpha_{f}$, $\alpha_{f} = \displaystyle \max_{m \in \mathcal{V}_{f}} (\alpha_{m}^{\ast})$, $\mathcal{V}_{f}$ is a set of victim MUEs that is affected by HeNB $f$. Each MUE asks for a different muted rate and then the muted rate for the HeNB can be calculated based on the required rate for the victim MUEs.

\subsection{Discussion}
\label{Distributed-Al}
Networks can have different architecture according to service providers and have different sizes of femtocells. According to the architecture, the algorithm can be operated in a different way. Also, in the case when the number of femtocells in a macrocell is large, due to computational complexity, a distributed algorithm can be more efficient while a centralized algorithm is more effective when the number of femtocells is small. Hence, we discuss about the distributed operations of the proposed algorithm.

For a distributed algorithm, Algorithms 1 and 2 are performed by the aggressor HeNBs autonomously. By exchanging information from the macro eNB, the aggressor HeNBs can autonomously decide to join or leave coalitions. In such coalitions, the aggressor HeNBs that affect the same VMUEs are grouped. To that end, while $\emph{Steps~1~\text{and}~2}$ are performed using the centralized algorithm, instead of using the coalition algorithm with the macro eNB in $\emph{Steps~3~\text{and}~4}$, the macro eNB sends the aggressor HeNBs the victim MUE information and then the aggressor HeNBs perform coalition formation on behalf of the macro eNB. All cells including the macro eNB and the HeNBs use the X2 interface in order to exchange signal information. For the distributed method, the proposed algorithm does $\emph{Steps~3~\text{and}~4}$  as follows:
\begin{itemize}
  \item \emph{Step 3}: After determining the victim macro users and their required muted rates, the macrocell sends the aggressor femtocell the information of the corresponding victim macro users. Then, the aggressor HeNB creates a list of victim macro users with their muted rated.

  \item \emph{Step 4}: The aggressor femtocells exchange lists via the X2 backhaul link in order to form coalitions if they affect in the same victim macro users. The aggressor femtocells decide whether to join or leave a coalition based on whether they have the same victim macro users in their list. The aggressor femtocell itself creates a particular coalition if the victim macro users are affected by only this one.
\end{itemize}

\section{Performance Evaluation}
\label{Evaluation}
In this section we use the Monte Carlo simulation in order to evaluate the performance of the macro users and femtocell users in LTE-A downlink. In each scenario, we derive the average muted rate required for the macro users, then the muted rates of the HeNBs are determined based on their affecting MUEs and the agreement in their coalitions.

For performance comparison, we consider two metrics: user throughput and outage probability. The user throughput is measured as the average amount of data received by users. Even though the overall throughput is high, individual QoS may not be satisfied. Hence, we define a user to be unsatisfied when the measured SINR of the user is lower than the minimum requirement, and  measure the ratio of the number of unsatisfied macro users to the total number of macro users, referred to as the outage probability.

\subsection{Simulation Environments}
\label{Eval-Envi}
Multi-users are considered, including multi-macro users and multi-femtocells in each scenario. The macro users are uniformly distributed within the coverage of macro eNB, and several HeNBs are also uniformly deployed surrounding the MUE.

Our proposed algorithm is compared with fixed ABSF, ABSF offseting~\cite{Kamel2013} and non-ABSF technique based algorithms. We use the following notations: Optimal ABSF, Fixed ABSF - I, Fixed ABSF - II, Fixed ABSF - III, and Non-ABSF  to represent our proposed algorithm, previous works with fixed ABSF ratios $ \displaystyle \frac{1}{10}, \displaystyle \frac{2}{10}, \displaystyle \frac{3}{10}$, and $0$, respectively. The non-ABSF technique means that if there is an available resource for the femtocell, the femtocell will transmit data to its own users without considering the presence of macro users. Consequently, the femtocell creates interference that affecting the nearby macro user performance.

In this paper, we assume the path loss model according to the urban deployment scenario~\cite{R4-092042}. The path loss is modeled at different types of links depending on the position and type of users. We consider different path loss models according to the environments: the downlink between serving the macro eNB, and the MUE and the downlink between the HeNB and the MUE.

The path loss between the serving  macro eNB and the outside MUE can be expressed in~Table~\ref{Tab-PathLoss}, where $D$ is the distance between the  macro eNB and the MUE in meters. The path loss between the aggressor HeNB and the MUE can also be expressed in~Table~\ref{Tab-PathLoss}, where $d$ is the distance between the HeNB and the MUE in meters.
\begin{table} [t]
\caption{Path loss~\cite{R4-092042}} 
\centering 
\begin{tabular}{l l} 
\hline 
Path Loss & Value (dB) \\ 
\hline 
Between serving  macro eNB and MUE                  & $L(dB) = 15.3 + 37.6 \log_{10}D$ \\
Between aggressor HeNB and MUE              & $L(dB) = 127 + 30 \log_{10} (d/1000)$  \\ %
\hline 
\end{tabular}
\label{Tab-PathLoss} 
\end{table}

After calculating path loss, the shadowing model is considered, and all links are taken into account for the shadowing model by adding log-normally distributed shadowing with standard deviation of 10 dB or 8 dB for links between the  macro eNB and the MUE, and between the HeNB and the MUE, respectively.

The system scenario assumptions and parameters are set up based on Monte-Carlo simulation~\cite{TR-36942}. The transmission power of the  macro eNB and HeNB are set to 46~dBm and 23~dBm, respectively. The antenna gain of the macro eNB and HeNB are also set to 14~dBi and 2.2~dBi, respectively. The parameter settings are summarized in~Table~\ref{parameter}. The requirement for the SINR threshold is either -6~dB or -4~dB according to the 3GPP standardization~\cite{yang2011a}, in this paper we assume the threshold SINR of MUE, $\gamma _{m}^{min}$, is set to -3~dB~\cite{Kamel2012}.
\begin{table}[t]
\caption{Parameter settings~\cite{TR-36942}} 
\centering
\begin{tabular}{l l} 
\hline 
Parameter & Values \\ 
\hline 
System bandwidth                    & 10 MHz \\ 
Duplex Technique                    & FDD mode \\ 
Frequency Reuse Scheme              & Reuse-1 \\ 
Macro eNB Transmission Power             & 46~dBm\\ %
Macro eNB Antenna Gain                   & 14 dBi\\ %
MUE Antenna Gain                    & 0 dBi\\ %
HeNB Antenna Gain                   & 2.2 dBi\\ %
Thermal Noise                       & -174 dBm/Hz \\
Noise Figure                        & 9~dB  \\
\hline 
\end{tabular}
\label{parameter} 
\end{table}

\subsection{Simplified Small Cell Network Scenario}
\label{Eval-Simp}
In this subsection we estimate the muted rate required for the macro user in a simplified scenario. The simplified scenario considers only one MUE served by  macro eNB and only one HeNB serves one HUE. The positions of MUE and HeNB are in the same geographical area. Two scenarios are considered here: center area and edge area scenarios. The center area can be defined as the interior of a circle in which the center point is the position of the  macro eNB and the radius of the circle is less than halft of the transmission range of the  macro eNB; in this scenario the radius is set to 200~meters. The edge area scenario is the rest of the  macro eNB coverage that does not contain the above-defined center area.

\begin{figure}[ht]
        \centering
         \includegraphics[scale=0.5]{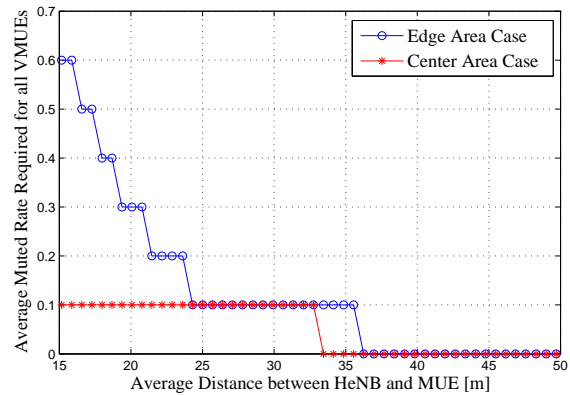}
        \caption{Average muted rate for the MUE in a simplified scenario}\label{SimplifiedScn}
\end{figure}

In Fig.~\ref{SimplifiedScn}, the average muted rate required for the MUE is reported in the center area scenario and the edge area scenario, respectively. The MUE in the center area is merely affected by the nearby HeNB, because if the MUE receives a strong desired signal from its serving  macro eNB, then the average muted rate is small. However, because the MUE located in the edge area receives both the weak desired signal from the macro eNB and the strong interfering signal from the nearby HeNB, then a higher muted rate is required than for the center area. The required muted rate for the MUE depends on both the relative distance between the  macro eNB and itself, and between the HeNB and itself. For a different MUE, the required muted rate is totally different; using the same muted rate for  all VMUEs is not a feasible solution.

\subsection{Static Multi-User Small Cell Network Scenario}
\label{Eval-Static}
In this scenario, we consider multi-users including MUEs, HeNBs, and HUEs deployed under macro eNB coverage. The multi-user scenario can be categorized into three types of network environments. In the first scenario, we assume that the number of HeNBs is larger than the number of MUEs (for instance, 60 HeNBs, 60 HUEs, and 40 MUEs are deployed). In the second scenario, the number of HeNBs is equal to the number of MUEs (for instance, 40 HeNBs, 40 HUEs, and 40 MUEs are considered). For the last scenario, the number of HeNBs is considered smaller than the number of MUEs; for example, there are 40 MUEs, 20 HeNBs, and 20 HUEs. For each type of femtocell networks, we will show the required blanking rate required for MUEs.
\begin{figure}[t]
  \centering
  \includegraphics[scale=0.5]{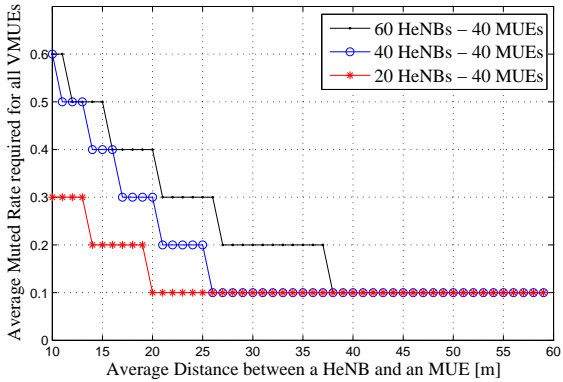}\\
  \caption{The average muted rate required by the victim MUEs}\label{All-MutedRate}
\end{figure}

In Fig.~\ref{All-MutedRate}, the average muted rate required for macro users is shown in three network environments. The user ratio is defined as the number of HeNBs divided by the number of MUEs. When the user ratio is larger than one, the required muted rate for the victim MUEs is higher than in other cases. As in the first scenario, the muted rate $\alpha_m$ is larger than that of the second and the third scenarios. It means that the muted rate $\alpha_m$ also depends on the number of HeNBs; the more HeNBs, the higher required muted rate $\alpha_m$.

\subsection{Dynamically Dense Small cell Network Scenario}
\label{Eval-Dynamically}
We study the impact of node mobility on the user performance in practical deployment. For the mobility model, we assume that macro users randomly choose the speed direction and the speed value, and change their speed direction and value at any time. The speed value is chosen randomly between the minimum (1~m/s) and the maximum value (20~m/s).

\begin{figure}[t]
  \centering
  \includegraphics[scale=0.5]{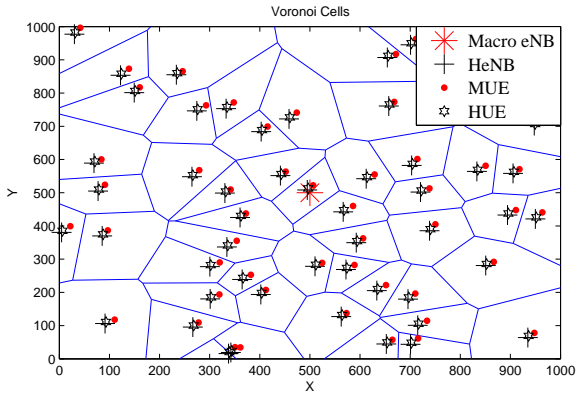}\\
  \caption{The Network Topology}\label{Voronoi_Topo}
\end{figure}

In Fig.~\ref{Voronoi_Topo}, an example of heterogeneous networks consisting of macro eNBs and HeNBs is shown. The numbers of the MUEs, HeNBs, and HUEs are set to 45. The locations of the cells and the user equipments are random;after initializing the locations, the macro users move according to the above mentioned mobility model.

\begin{figure}[t]
  \centering
  \includegraphics[scale=0.5]{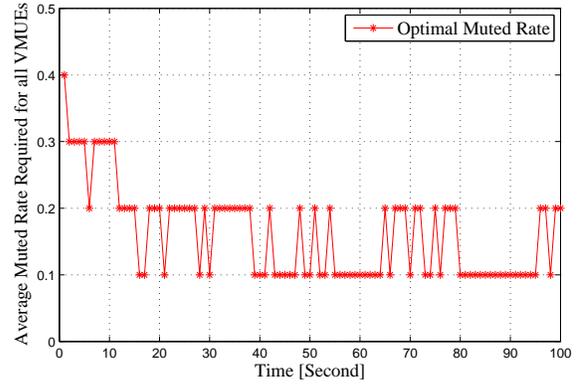}\\
  \caption{The muted rate required by victim MUEs}\label{M-MutedRate}
\end{figure}

We derive the muted rate required for the MUEs as shown in~Fig.~\ref{M-MutedRate}. The number of ABSFs required for the MUEs can be calculated by multiplying the muted rate and the number of sub-frames in each radio frame. The results show that the number of ABSFs required varies with time. Hence, if the ABSF ratio is set to a fixed value, the demand of the MUE may be higher or lower than the fixed value. A dynamically optimal muted rate is the best solution for eICIC technique in the time domain.

\begin{figure}[t]
        \centering
        \includegraphics[scale=0.5]{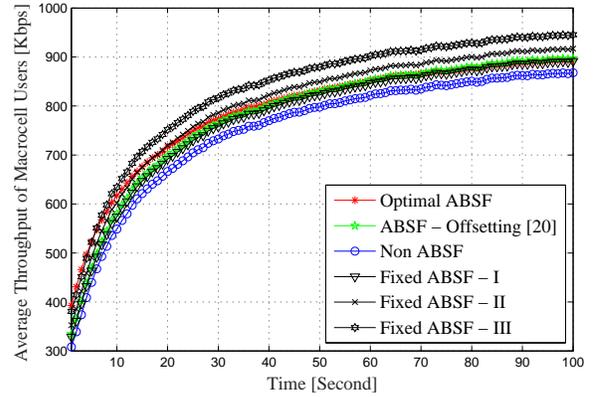}
        \caption{Average Macrocell Users Throughput [kbps]}
        \label{fig:M-Thr-Macro}
\end{figure}

We compare our proposed algorithm with previous work in terms of user throughput. The eICIC technique can help improve the performance of the MUEs; as can be seen in Fig.~\ref{fig:M-Thr-Macro}, the algorithms using the ABSF technique outperform the algorithm without the ABSF technique. Our proposed algorithm can adjust the ABSF rate according to the macro user requirements, while other algorithms with the same muted rate can offer more or less radio resources for the macro users in order to maintain MUE performance. As compared to~\cite{Kamel2013}, the muted rate is derived in order to maximize the total network utility, which depends mainly on the total number of VMUEs and number of VMUEs per each HeNB. However, in this considered scenario, the number of VMUEs per each HeNB is small (mostly 1 VMUE per each HeNB), then the reduced blacking rate is 1/8. On the other hand, the number of aggressor HeNBs per VMUEs is normally higher than 1, then the aggressor HeNBs are better to cooperate in order to mitigate the cross-interference efficiently. Due to the node movement, in some cases, the number of VMUEs around HeNBs is larger, then the total network utility of proposed algorithm in~\cite{Kamel2013} is slightly higher than that our proposed algorithm.

\begin{figure}[t]
        \centering
      \includegraphics[scale=0.5]{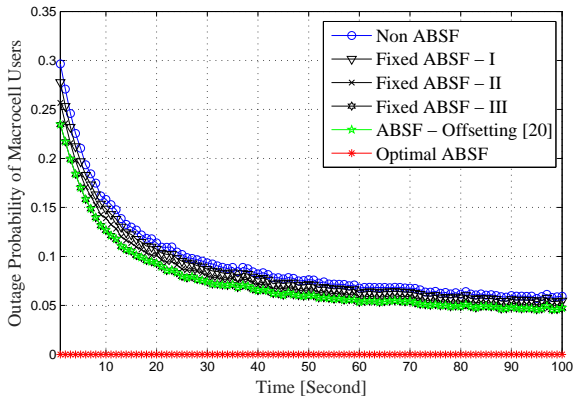}
      \caption{Outage Probability of Macro Users}
      \label{fig:M-Outage_M}
\end{figure}

Figure~\ref{fig:M-Outage_M} shows the outage probability of the macro users. The outage probability is defined as the ratio of the number of unsatisfied macro users to the total number of macro users. Our proposed algorithm is developed in order to satisfy the QoS of all macro users so that there are no macro users with a SINR less than the SINR threshold. In contrast, the previous work could not guarantee the QoS of macro users in all cases, even with a high muted rate. Hence, our proposed algorithm outperforms the previous work.

\begin{figure}[t]
        \centering
        \includegraphics[scale=0.5]{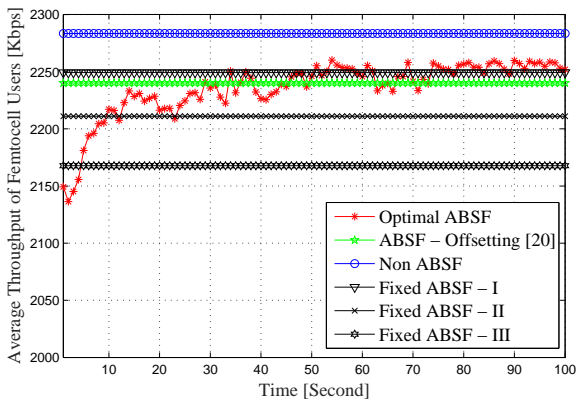}
        \caption{Average Femtocell User Throughput [kbps]}
        \label{fig:M-Thr-Femto}
\end{figure}

\begin{figure}[t]
      \centering
      \includegraphics[scale=0.5]{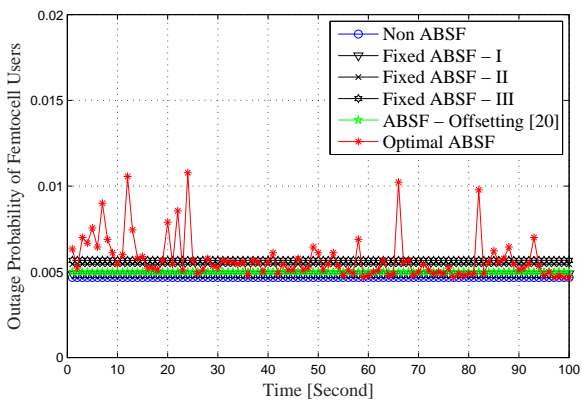}
      \caption{Outage Probability of Femtocell Users}
      \label{fig:M-Outage-Femto}
\end{figure}

Figures~\ref{fig:M-Thr-Femto} and~\ref{fig:M-Outage-Femto} show the throughput and the outage probability of small cell users, respectively. As compared with previous algorithms, our proposed algorithm can adapt to the network environment, the small cell performance of our proposed algorithm is better than that of the previous works with a high muted rate in terms of user throughput. The small cell outage probability of all algorithms is very small, almost zero, suggesting that in this case the performance of small users is still acceptable.

\section{Conclusion}
\label{Conclusion}
This paper proposes a dynamically optimal ABSF eICIC framework in order to mitigate the impact of cross-layer interference in HetNets. Unlike previous works, the number of ABSFs depends on network environments such as the numbers of macro users and small cells and QoS requirements. In this paper, the number of ABSFs is derived based on the QoS of each macro user MUE, and then based on the required muted rate for each MUE we can set a dynamically optimal blank rate for each HeNB. Obviously, since each HeNB has unique characteristics, each HeNB has to decide its own muted rate. Due to the mutual interference among HeNBs, HeNBs should cooperate to mute their radio resource in order to protect the MUEs effectively. We also propose a coalition algorithm to help HeNBs use the ABSF framework efficiently. Simulations showed that our proposed algorithm outperforms the previous work in various environments.

\bibliographystyle{ieicetr}
\bibliography{FemtocellRef}


\profile[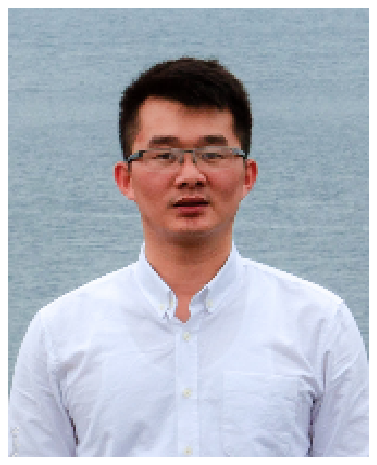]{Trung Kien Vu}{received B.Eng. degree in the School of Electronics and Telecommunications, Hanoi University of Science and Technology, Hanoi, Vietnam in 2012, and the M.Sc. degree in electrical engineering in the School of Electrical Engineering, University of Ulsan, Korea in 2014. He is currently pursuing the D.Sc. degree at Centre for Wireless Communications (CWC), University of Oulu, Finland. He is also a member of the research stuff of CWC. His research interests include heterogeneous cellular networks, mobile ad-hoc networks.}
\profile[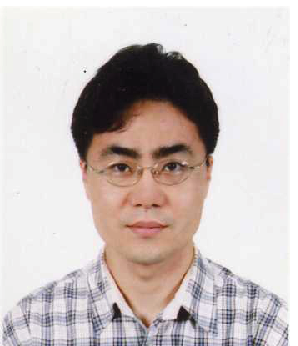]{Sungoh Kwon}{received the B.S. and M.S. degrees in electrical engineering from KAIST, Daejeon, Korea, and the Ph.D. degree in electrical and computer engineering from Purdue University, West Lafayette, IN, in 1994, 1996, and 2007, respectively. From 1996 to 2001, he was a research staff member with Shinsegi Telecomm Inc., Seoul, Korea. From 2007 to 2010, he developed LTE schedulers as a principal engineer in Samsung Electronics Company, Ltd., Korea. He has joined to University of Ulsan as an assistant professor since 2010 and is now an associate professor. His research interests are in wireless communication networks.}
\profile[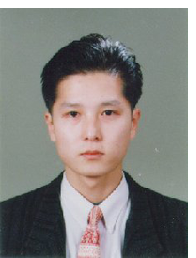]{Sangchul Oh}{received B.S. and M.S. degrees from the Kwangwoon University, Seoul, Korea, in 1995 and 1997, respectively, all in Electronics and Telecommunications Engineering. He worked as a researcher in the Daewoo Telecommunications from 1997 to 2000. He is currently a principal researcher at ETRI. His research interests are in LTE and LTE-Advanced System.}

\end{document}